\documentclass[twocolumn,amsmath,amsfonts,showpacs,floatfix]{revtex4}
\usepackage{graphicx}
\usepackage{epstopdf}
\usepackage{tikz}
\usepackage{float}

\newcommand{\ket}[1]{$|#1\rangle$}

\begin{document}
\title{Read-out and Dynamics of the Qubit Built on Three Quantum Dots}

\author{Jakub {\L}uczak and Bogdan R. Bu{\l}ka}
\affiliation{Institute of Molecular Physics, Polish Academy of Sciences, ul. M. Smoluchowskiego 17, 60-179
Pozna{\'n}, Poland}

\date{\today}

\begin{abstract}
We present a model of a qubit built of a three coherently coupled quantum dots with three spins in a triangular geometry. The qubit
states are encoded in the doublet subspace and they are controlled by a gate voltage, which breaks the triangular
symmetry of the system. We show how to prepare the qubit and to perform one qubit operations. A new type of the current blockade effect will be discussed. The blockade is related with an asymmetry of transfer rates from the electrodes to different doublet states and is used to read-out of the dynamics of the qubit state. Our research also presents analysis of the Rabi oscillations, decoherence and leakage processes in the doublets subspace.
\end{abstract}


\maketitle

\section{Introduction}

 A quantum computer will allow to perform some algorithms much faster than in classical computers e.g. Shor
algorithm for the
factorization the numbers \cite{nielsen}. The basic elements in the quantum computation are qubits and quantum logical gates, which allow to construct any
circuit to quantum algorithms. The good candidates to realization of qubits are semiconductor quantum dots with controlled electron numbers. The qubit state can
be encoded using an electron charge or, which is also promising, an electron spin \cite{loss}. The spin qubits are characterized by longer decoherence times necessary in the quantum computation \cite{vrijen}. However to prepare that qubit one needs to apply a
magnetic field and removed the degeneracy between spin up and down. The manipulation of the qubit can be done by electron spin resonance and the read-out via
currents in spin-polarized leads \cite{engel}. Another concept to encode the qubit is based on the singlet-triplet states in a double quantum dot (DQD). In
this case the magnetic field is not necessary and the qubit preparation is performed by electrical control of the exchange interactions
\cite{barthel,petta}. The qubit states can be controlled by e.g. an external magnetic field \cite{koppens}, spin-orbit \cite{golovach} or
hyperfine interaction \cite{bluhm, 2qdtheory}.
For the read-out of the qubit
state one can use current measurement and the effect of Pauli spin blockade \cite{liu}. In the Pauli blockade regime the current flows only for the singlet,
which gives information about the qubit states.

DiVincenzo {\it et al} \cite{vinzenzo} suggested to build the qubit in more complex system, namely in three coherently couplet quantum dots (TQD). The qubit
states are encoded in the doublet subspace and can be controlled by exchange interactions. This subspace was pointed as a decoherence-free subspace (DFS)
\cite{dfs}, which is immune to decoherence processes.
Another advantage of this proposal is the purely electrical control of the exchange interactions by gate potentials which act locally and provide much faster operations.
In the TQD system, in the contrast to the DQD qubit, one can modify more than one exchange interaction between the spins and perform
full unitary rotation of the qubit states \cite{nielsen}. The three spin qubit has also more complicated energy spectrum which provides operations on more states in contrast to the two spin system.
Recently experimental efforts were undertaken \cite{laird,aers,gaudreau,amaha} to get
coherent spin manipulations in a linear TQD system according to the scheme proposed by DiVincenzo {\it et al}  \cite{vinzenzo}.
The initialization, coherent exchange and decoherence of the qubit states were shown in the doublet \cite{laird,aers} and doublet-quadruple subspace
\cite{gaudreau}. The read-out of the qubit state was performed, like in DQD, by means of the Pauli blockade \cite{laird,aers,gaudreau}. Amaha {\it et al.}
\cite{amaha} observed a quadruplet blockade effect which is based on reducing leakage current from quadruplet to triplet states in the presence of magnetic
field. Shi {\it et al.} \cite{shi} showed that DiVincenzo's proposal can be realized on double quantum dots with many levels and three spin system controlled by gate potentials.

In this paper we demonstrate that TQD in a triangular geometry can work as a qubit. This kind of TQD was already fabricated experimentally by local
anodic oxidation with the atomic force microscope \cite{rogge} and the electron-beam lithography \cite{seo}. In the triangular TQD qubit exchange interactions between all spins are always on and very important is symmetry of the system.
Trif et al. \cite{trif, trif2010} and Tsukerblat \cite{tsukerblat} studied an influence of the electric field on the symmetry of triangular molecular magnets and spin configurations in the presence of a spin-orbit interaction.
DiVincenzo's scheme to encode the qubit in triangular TQD was considered by Hawrylak and Korkusinski \cite{hawrylak} where one of the exchange coupling was modified by gate potential. Recently Georgeot and Mila \cite{georgeot} suggested to build the qubit on two opposite chiral states generated by a magnetic flux penetrating the triangular TQD. One can use also a special configuration of magnetic fields (one in-plane and perpendicular to the TQD system) to encode a qubit in chirality states \cite{hsieh}. Recent progres in theory and experiment with TQD system was reported in \cite{hsiehRep}.

Our origin idea is to use the fully electrical control of the symmetry of TQD to encode and manipulate the qubit in the doublet subspace. The doublets are vulnerable to change the symmetry of TQD, which will be use to
prepare and manipulate the qubit (sec. \ref{preparation}). The crucial aspect in quantum computations is to read-out the qubit states. Here we propose a new detection method, namely, a doublet blockade effect which manifests itself
in currents for a special configuration of the local potential gates. We show (sec. \ref{detection}) that the doublet blockade is related with an asymmetry of a tunnel rates from source and drain electrodes to TQD and the inter-channel Coulomb blockade. The method is fully compatible with purely electrical manipulations of the qubit. Next we present studies of dynamics of the qubit and demonstrate the coherent and Rabi
oscillations (sec. \ref{dynamics}). The studies take into account relaxation and decoherence processes due to coupling with the electrodes
as well as leakage from the doublet subspace in the measurement of current flowing through the system. We derive characteristic times which describe all relaxation processes.  Our model is general and can be used for a qubits encoded also in the linear TQD, which is a one of the cases of broken symmetry in the triangular TQD.

\section{Model}\label{modelTQD}

\begin{figure}[ht]
\includegraphics[width=0.25\textwidth,clip]{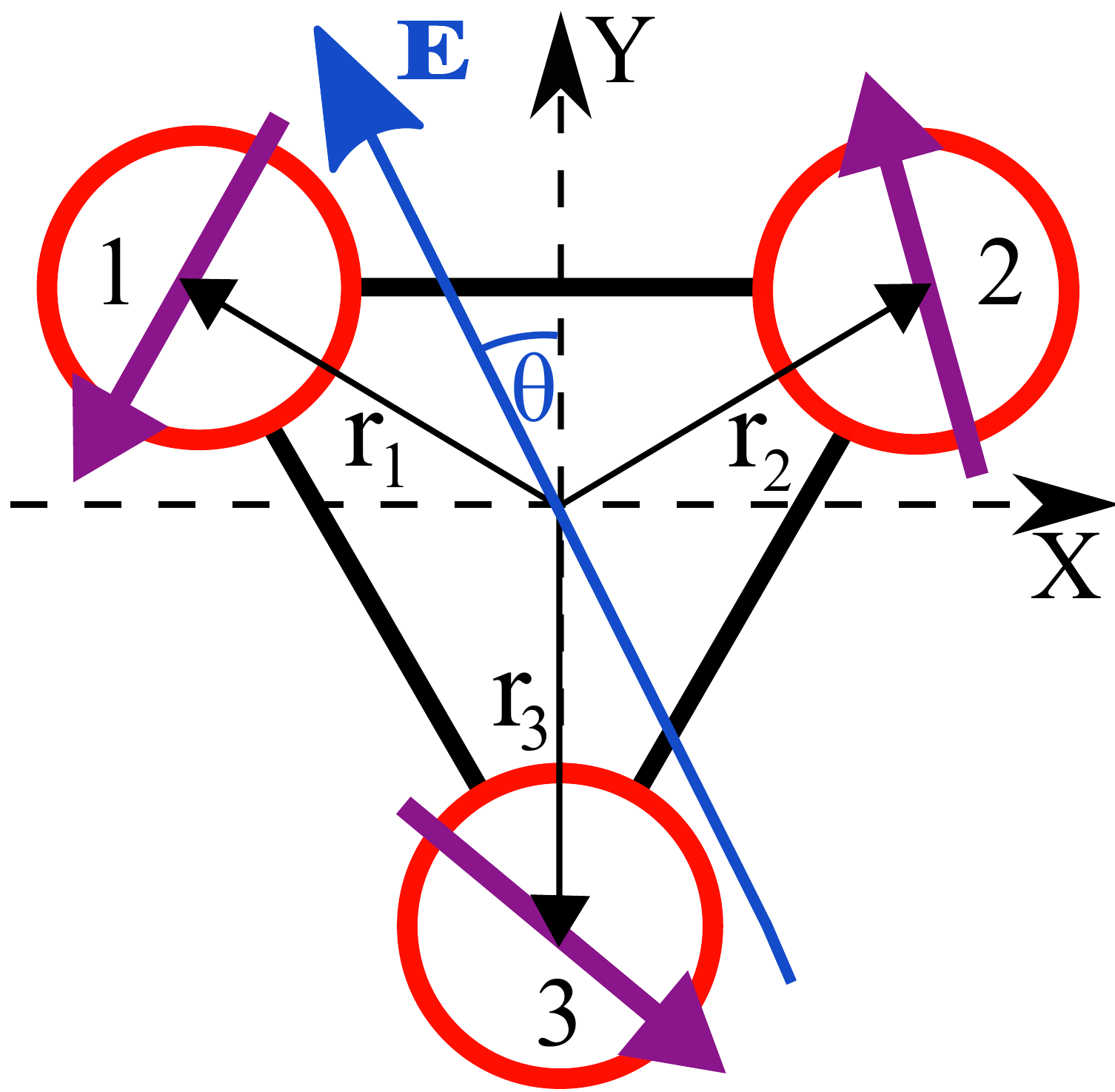}
\caption{The model of triangular molecule placed in the effective electric field $\mathbf{E}$.}
\label{fig1}
\end{figure}

Our system is a triangular artificial molecule built of three coherently coupled quantum dots with a single electron spin
on each dot (see. Fig.\ref{fig1}). Interactions between the spins are described by an effective Heisenberg Hamiltonian
\begin{eqnarray}\label{heisenberg}
\hat{H} = \sum_i J_{i,i+1}(\mathbf{S}_i\cdot\mathbf{S}_{i+1}- \frac{1}{4})-g \mu_B B_z \sum_i S_{z,i}\,,
\end{eqnarray}
where the Zeeman term is included to show splitting by an external magnetic field $B_z$ ($\mu_B$ is the Bohr
magneton, \emph{g} is the electron g-factor) and $J_{i,i+1}$ is an exchange interaction between electrons on sites $i$ and $i+1$.

The exchange parameter can be calculated by Heitler-London and Hund-Mulliken method. For a defined confinement potential one can find the parameter as a function of the interdot distance, the potential barrier and the magnetic field  \cite{burkard, cywinski}.

For the system with three spins there are two subspaces, one of them is a quadruplet with the total spin $S=3/2$ and
$S_z=\pm1/2,\pm3/2$. The quadruplet states are given by:
\begin{eqnarray}
|Q^{1/2}\rangle=\frac{1}{\sqrt{3}}(|\uparrow_1\uparrow_2\downarrow_3\rangle
+|\uparrow_1\downarrow_2\uparrow_3\rangle+|\downarrow_1\uparrow_2\uparrow_3\rangle),
\\
|Q^{3/2}\rangle=|\uparrow_1\uparrow_2\uparrow_3\rangle,
\end{eqnarray}
and similar functions for opposite spin orientations. Energies of these states are
$E_{Q}^{S_z}=- g \mu_B B_zS_z$.
The second subspace is formed by doublet states with $S=1/2$ and $S_z=\pm1/2$. The doublet state for $S_z=1/2$ can be expressed as:
\begin{eqnarray}\label{doublet}
|\Psi_D^{1/2}\rangle=\alpha_{D_1}|D_1^{1/2}\rangle+\alpha_{D_2}|D_2^{1/2}\rangle,
\end{eqnarray}
where
\begin{eqnarray}\label{d1}
|D_1^{1/2}\rangle&=&\frac{1}{\sqrt{2}}(|\uparrow_1\uparrow_2\downarrow_3\rangle-|\uparrow_1\downarrow_2\uparrow_3\rangle) \equiv
|\uparrow_1\rangle|S_{23}\rangle\\ \label{d2}
|D_2^{1/2}\rangle
&=&\frac{1}{\sqrt{6}}(|\uparrow_1\uparrow_2\downarrow_3\rangle+|\uparrow_1\downarrow_2\uparrow_3\rangle -2|\downarrow_1\uparrow_2\uparrow_3\rangle)
\nonumber \\
&\equiv&\frac{1}{\sqrt{3}}|\uparrow_1\rangle|T_{23}^0\rangle-\sqrt{\frac{2}{3}}|\downarrow_1\rangle |T_{23}^{1} \rangle,
\end{eqnarray}
$|S_{23}\rangle$ and $|T_{23}^{S_z}\rangle$ denote a singlet and triplet state on the $\{23\}$  bond, respectively. Here we assume large Coulomb intradot interactions and ignore double electron occupancy.
  In the doublet subspace (\ref{d1})-(\ref{d2}) one can express the Hamiltonian (\ref{heisenberg}) as
\begin{eqnarray}\label{matrix2}
\hat{H_\sigma}=-\frac{1}{2}(3J+g \mu_B B_z)\mathbf{1}+\frac{\delta}{2}\sigma_z+\frac{\gamma}{2}\sigma_x
\end{eqnarray}
using the Pauli matrix representation. The parameters are given by
\begin{eqnarray}\label{jjj}
J &=& \frac{1}{3}(J_{12}+J_{23}+J_{31}), \\ \label{delta}
\delta &=& \frac{1}{2}(J_{12}+J_{31}-2J_{23}), \\ \label{gamma}
\gamma &=& \frac{\sqrt{3}}{2}(J_{12}-J_{31}).
\end{eqnarray}
The eigenvalues of (\ref{matrix2}) are:
\begin{eqnarray}
E_{D}^\pm= -\frac{3}{2}J-\frac{g \mu_B B_z}{2} \pm \frac{\Delta}{2},
\end{eqnarray}
where $\Delta=\sqrt{\gamma^2+\delta^2}$ is the doublet splitting and $J$ describes the energy differences between the doublet and the quadruplet subspace. Two other parameters $\delta$ and $\gamma$ can be interpreted as an effective magnetic field in the {\it z} and {\it x} direction, respectively \cite{gimenez}.

In GaAs/AlGaAs quantum dots the exchange interaction $J$ is estimated in the range $ 0.02 \div 0.3$ meV \cite{busl,taylor} and in a molecular magnet as $0.43$ meV \cite{trif}. The doublet splitting for a linear TQD is the order $\Delta \sim 0.1\mu \div 1$ $\mu$eV \cite{gaudreau,mehl}. This parameter can be even larger $\Delta\sim 21.6$ $\mu$eV  in Si/SiGe quantum dots \cite{shi2013}.

In this paper we assume that the exchange couplings $J_{i,i+1}$ can be manipulated by local potential gates $V^{gate}_{i,i+1}$, which change potential barriers and modify electron hopping as well as local covalency between the quantum dots.
The exchange coupling can be expressed in the linear approximation as $J_{i,i+1}=J+j_V V^{gate}_{i,i+1}$, where $j_V$ describes sensitivity of the exchange coupling to the gate voltage.
For our analysis of the symmetry breaking in TQD, it is more suitable to parameterize the gate potentials as $V^{gate}_{i,i+1}= V_0+v \cos\left[\theta+(i-1)2\pi/3\right]$ with  $V_0\equiv0$, some amplitude $v$ and angle $\theta$.
This parametrization corresponds to influence of an effective electric field ${\bf E}$ on the bond polarization and covalency.
For a small  value of {\bf E} the exchange couplings can be expressed as
\begin{eqnarray}\label{exchangeinE}
J_{i,i+1}=J+g_E\;\cos\left[\theta+(i-1)\frac{2\pi}{3}\right],
\end{eqnarray}
where $g_E= j_E e|{\bf E}||{\bf r}_1-{\bf r}_2|$, $j_E$ is a parameter describing sensitivity of the exchange coupling to the electric field, $e$ - the elementary electron charge, ${\bf r}_i$ - the vector showing the position of the $i$-th quantum dot,
$\theta$ is the angle between ${\bf E}$ and the axis $Y$ (see Fig.\ref{fig1}). A similar relation was obtained for the triangular molecule in the electric field which changed chirality of the spin system \cite{trif,trif2010}.
Let us stress that because the electric field is taken as the small parameter, single electron occupancy of each dot is conserved and the ground state is always the doublet.

In the TQD system  one can also consider superexchange processes through excited double occupied states. Applying local potential gates to the quantum dots one can shift their energy levels and modify the superexchange couplings
\cite{bulka}. Because a parameter of inter-dot electron hopping is relatively small with respect to a intra-dot Coulomb interaction, the modifications of the superexchange couplings are very small and will not be discussed in the paper.

\section{Qubit preparation and manipulation}\label{preparation}

Let us consider how to encode the qubit in the doublet subspace with the spin $S_z=1/2$, expressed by \ket{\Psi_D^{1/2}}. This state is isolated from the doublet \ket{\Psi_D^{-1/2}} and the quadruplets states for a moderate magnetic field, $\Delta<g \mu_B B_z<(E_Q^{3/2}-E_D^+)$. The encoded qubit states $|0\rangle$ and $|1\rangle$ correspond to the doublets \ket{D_1} and \ket{D_2}, Eq.(\ref{d1}) and (\ref{d2}) (the spin index $1/2$ is omitted to simplify the notation).
In the further considerations the hyperfine and spin-orbit interactions are ignored.

\begin{figure}[ht]
\centering
\includegraphics[width=0.3\textwidth, clip]{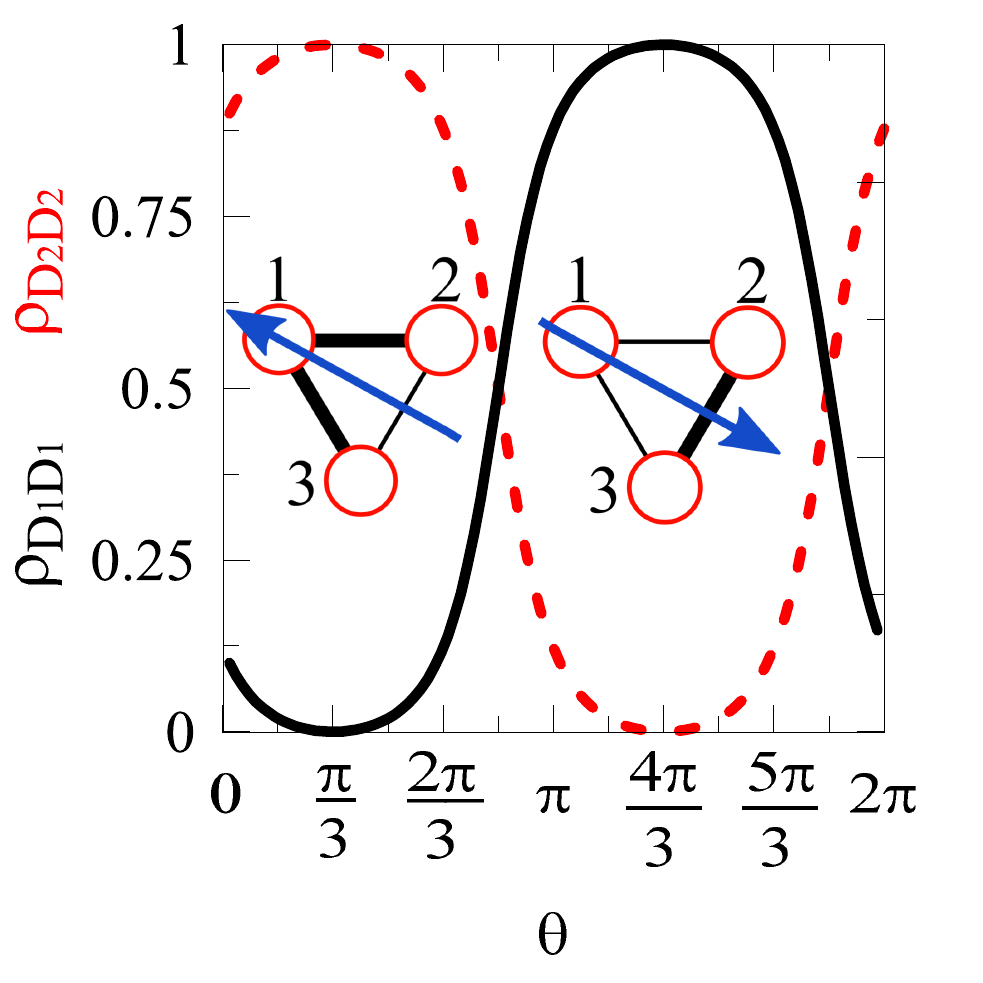}
\caption{Density matrix elements $\rho_{D_1D_1}$ (solid black curve) and $\rho_{D_2D_2}$ (dashed red curve) vs $\theta$ for the ground
state at a moderate value of the electric field $g_E=0.2, J=1$. The left and right inserts shows direction of the electric field when the qubit is prepared in the states \ket{D_2} and \ket{D_1}, respectively. These two opposite directions of the electric field corresponds to a minimal (thin line) and maximal (thick line) bond polarization between the dots 2 and 3.}
\label{eigenvectors}
\end{figure}

The qubit is prepared by a proper orientation $\theta$ of the effective electric field $\bf E$ which changes the symmetry of the system. Fig. \ref{eigenvectors}
presents density matrix elements $\rho_{D_1D_1}$ and $\rho_{D_2D_2}$ as a function of $\theta$. One can see that the qubit is prepared in the state \ket{D_1} for  $\theta=4\pi/3$ (the electric field is oriented from the quantum dot 1).
For this symmetry the exchange parameters are $J_{12}=J_{31}<J_{23}$ and from eq. (\ref{delta}) and (\ref{gamma}) one gets $\delta<0$ and the mixing
between the doublets $\gamma=0$.
For $\theta=\pi/3$ the electric field points to the quantum dot 1 and $J_{12}=J_{31}>J_{23}$, $\delta>0$ and $\gamma=0$. In this case the qubit is prepared in the state $|D_2\rangle$. We would like to emphasize that in the triangular TQD the both qubit states are equivalent and can be easily achieved only by change the symmetry of the system. This is the main advantage in comparison with the linear TQD where the qubit can be prepare usually only in one of the doublet state.

Now we show how one can perform one qubit operations by means of the electric field. After preparation of the qubit in one of the states \ket{D_1} or \ket{D_2} we change rapidly the angle $\theta$ to perform a dynamic rotation of the qubit state.
The qubit dynamics is described by the time-dependent Schr\"{o}dinger equation with the Hamiltonian (\ref{matrix2}). We can show two basic quantum gates. For $\delta=0$ the pseudo-spin rotates around the x-axis on the Bloch sphere, which is the Pauli-x quantum gate and the solution of the Schr\"{o}dinger is given by
\begin{eqnarray}\label{h2}
|\Psi_D(t)\rangle=\exp\left[\frac{i}{2}(3J+g\mu_B B_z) t \right]U_x |\Psi_D(0)\rangle\;.
\end{eqnarray}
Here $U_{x}=\exp\left[-i \gamma \sigma_x t/2\right]$ is an unitary operator of rotation around the x-axis. Second quantum gate we get for $\gamma=0$ with the solution given by (\ref{h2}) but now instead $U_x$ we have $U_z=\exp\left[-i \delta \sigma_z t/2\right]$ which is an unitary operator of rotation around the z-axis. These two rotations can be use to get to any point on the Bloch sphere. It is clearly seen that by modification of the parameters $\gamma$ and $\delta$ one can get full control of the qubit operations (see also \cite{weinstein}).

\section{Detection - doublet blockade} \label{detection}

In the linear TQD the read-out of the qubit state is possible due to charge-spin conversion in the regime of the Pauli spin blockade. A detuning voltage is applied between two outermost dots, which drives the system from single occupied configuration (1,1,1) to double occupied e.g. (2,0,1). This transfer is possible for an electron with the opposite spin orientation and can be detected by a quantum point contact (QPC) \cite{gaudreau,laird}.

In this section we would like to show a new method to read-out the qubit state which is based on a measurement of currents flowing through the system. The detection is compatible with electrical control of qubit state and the charge-spin conversion is not necessary. We assume that TQD is coupled by tunnel
junctions to the electrodes, where the first and the second dot are connected to the left and right electrode, respectively. Application of this method in an experimental setup is a similar technical complexity as QPC. The electron transport through the tunnel junctions is studied within the sequential tunneling regime. Transfer rates from the left (L) and the right (R)
electrode to TQD are given by:
\begin{eqnarray}\label{transferRates}
\Gamma^{L(R)+}_{\nu_2\rightarrow\nu_3}=\frac{\Gamma}{\hbar}|\langle \nu_3| c_{1(2),\sigma}^\dagger |\nu_2\rangle|^2
f(E_{\nu_3}-E_{\nu_2}-\mu_{L(R)}).
\end{eqnarray}
Here we assume that both tunnel barriers are characterized by the same parameter $\Gamma$, the reduced Planck constant is taken $\hbar=1$,
 \ket{\nu_2} and \ket{\nu_3} denote the states with two and three electrons with the corresponding energies $E_{\nu_2}$ and $E_{\nu_3}$,
$c_{1(2),\sigma}^\dagger$ is an electron creation operator on the dot 1 (2) with spin $\sigma=\pm 1/2$. \emph{f}
denotes the Fermi distribution function, the electrochemical potentials in the left and the right electrode are $\mu_L=E_F$ and $\mu_{R}=E_F+eV$, where $E_F$ is
the Fermi energy and $V$ is an applied bias voltage. By
analogy one can define transfer rates $\Gamma^{L(R)-}_{\nu_3\rightarrow\nu_2}$ from TQD to the electrodes.
We confine our considerations to a voltage window with transitions between
the states with three and two electrons, but a similar situation one can expect for transitions between three and
four electron states. Two electron states \ket{\nu_2} can be either the singlet \ket{S} or triplet \ket{T^{S_z}}.
For a high intra-dot Coulomb interaction one can neglect double occupied states and confine considerations to the states with single electron occupancy only. The
singlet can be then
expressed as a linear superposition:
$|S\rangle=\alpha^S_{12}|S_{12}\rangle +\alpha^S_{23}|S_{23}\rangle +\alpha^S_{31}|S_{31}\rangle$, where
$|S_{ij}\rangle=(|\uparrow_i \downarrow_j \rangle-|\downarrow_i \uparrow_j \rangle)/\sqrt{2}$ denotes the singlet on
the $\{i,j\}$ pair of dots.
Calculating the elements of the transfer matrices one can find net transfer rates between the doublet \ket{D_1^{\pm1/2}}, \ket{D_2^{\pm1/2}} and the singlet
\ket{S}:
\begin{eqnarray}\label{DtoS}
|\langle D_1^{\pm1/2}|c_{1\sigma}^\dagger|S\rangle|^2&=&|\alpha_{23}^S|^2,\\ \label{sd2eq}
|\langle D_2^{\pm1/2}|c_{1\sigma}^\dagger|S\rangle|^2&=&0,\\
|\langle D_1^{\pm1/2}|c_{2\sigma}^\dagger|S\rangle|^2&=&\frac{1}{2}|\alpha_{31}^S|^2,\\ \label{DtoS2}
|\langle D_2^{\pm1/2}|c_{2\sigma}^\dagger|S\rangle|^2&=&\frac{3}{4}|\alpha_{31}^S|^2.
\end{eqnarray}
For symmetry reasons there are no transfers between \ket{S} and \ket{Q^{S_z}}.
If we express the triplet state in the form $|T^{S_z}\rangle=\alpha^{T}_{12}|T^{S_z}_{12}\rangle +
\alpha^{T}_{23}|T^{S_z}_{23}\rangle +\alpha^{T}_{31}|T^{S_z}_{31}\rangle$, with
 $|T^1_{ij}\rangle=|\uparrow_i \uparrow_j\rangle$, $|T^0_{ij}\rangle=(|\uparrow_i \downarrow_j\rangle+|\downarrow_i
 \uparrow_j \rangle)/\sqrt{2}$ and  $|T^{-1}_{ij}\rangle=|\downarrow_i \downarrow_j\rangle$, then
the corresponding transfer elements are:
\begin{flalign}\label{DtoT1}
|\langle D_1^{\pm1/2}|c_{1,\overline{\sigma}}^\dagger|T^{\pm 1}\rangle|^2 & =|\langle
D_1^{\pm1/2}|c_{1,\sigma}^\dagger|T^{0}\rangle|^2=0,\\
|\langle D_2^{\pm1/2}|c_{1,\overline{\sigma}}^\dagger|T^{\pm 1}\rangle|^2 & =2|\langle
D_2^{\pm1/2}|c_{1\sigma}^\dagger|T^{0}\rangle|^2=
\frac{2}{3}|\alpha_{23}^T|^2, \\
|\langle D_1^{\pm1/2}|c_{2,\overline{\sigma}}^\dagger|T^{\pm 1}\rangle|^2 & =2|\langle
D_1^{\pm1/2}|c_{2,\sigma}^\dagger|T^{0}\rangle|^2=\frac{1}{2}|\alpha_{31}^T|^2, \\
\label{DtoT4}
|\langle D_2^{\pm1/2}|c_{2,\overline{\sigma}}^\dagger|T^{\pm 1}\rangle|^2 & =2|\langle
D_2^{\pm1/2}|c_{2\sigma}^\dagger|T^{0}\rangle|^2=
\frac{1}{6}|\alpha_{31}^T|^2,
\end{flalign}
and for quadruplets

\begin{eqnarray}
|\langle Q^{\pm3/2}|c_{1\sigma}^\dagger|T^{\pm1}\rangle|^2 = (3/2)|\langle
Q^{\pm1/2}|c_{1\sigma}^\dagger|T^{0}\rangle|^2= \nonumber\\ 3 |\langle Q^{\pm1/2}|c_{1,\overline{\sigma}}^\dagger|T^{\pm1}\rangle|^2  =
|\alpha_{23}^T|^2,\\
\label{QtoT}
|\langle Q^{\pm3/2}|c_{2\sigma}^\dagger|T^{\pm1}\rangle|^2= (3/2)|\langle
Q^{\pm1/2}|c_{2\sigma}^\dagger|T^{0}\rangle|^2= \nonumber\\ 3 |\langle Q^{\pm1/2}|c_{2,\overline{\sigma}}^\dagger|T^{\pm1}\rangle|^2=
|\alpha_{31}^T|^2.
\end{eqnarray}

The Hamiltonian in the singlet subspace $\{$\ket{S_{ij}}$\}$ is
\begin{eqnarray}\label{HS}
H_S=\left[
\begin{array}{ccc}
\tilde{\epsilon}_{12} & t_{23} & t_{31} \\
        t_{12} & \tilde{\epsilon}_{23} & t_{23} \\
        t_{31} & t_{23} & \tilde{\epsilon}_{31} \\
      \end{array}\right],
\end{eqnarray}
whereas for triplets $\{$\ket{T^{S_z}_{ij}}$\}$
\begin{eqnarray}\label{HT}
H_T=\left[
\begin{array}{ccc}
\tilde{\epsilon}_{12} & -t_{23} & -t_{31} \\
        -t_{12} & \tilde{\epsilon}_{23} & -t_{23} \\
        -t_{31} & -t_{23} & \tilde{\epsilon}_{31} \\
      \end{array}\right].
\end{eqnarray}
Here, $\tilde{\epsilon}_{i,i+1}=\varepsilon_{i}+\varepsilon_{i+1}+U_{i,i+1}$ denotes a local energy of two electrons on the
$\{i,i+1\}$ pair (in calculations we take $\varepsilon_{i}=0$) including an inter-dot coulomb interaction $U_{i,i+1}$. Here the hopping parameter is taken $t_{i,i+1}=t_0+t_E\,e|{\bf E}||{\bf r}_1-{\bf r}_2|\cos\left[\theta+2(i-1)\pi/3\right]$.
The difference between $H_T$ and $H_S$ is the sign in the off-diagonal elements, which makes difference in the
spectrum. For $t_{i,i+1}<0$ the ground state is singlet, whereas triplet becomes the ground state for $t_{i,i+1}>0$. The
ground state never can be a dark state, neither singlet nor triplet, the coefficients $\alpha^{S,T}_{i,i+1}\neq 0$ (see \cite{poltl} for
more details).

From Eq.(\ref{DtoS})-(\ref{DtoT4}) one can see that transfer rates from doublets are asymmetric.
An electron can tunnel from the right electrode to the both states \ket{D_1} and \ket{D_2}, but it can be transferred further to the left electrode through one
doublet only. In such the situation one can expect the inter-channel Coulomb blockade effect. If an electron is captured in one of the doublet state, it blocks
(due
to Coulomb interaction) flow  of electrons through the other state.
For transport from the singlet state the
electron can be captured at \ket{D_2} which results the current blockade through \ket{D_1}. In transport through the triplet the
role of the doublets is reversed, transport through \ket{D_2} is blocked by an electron captured at \ket{D_1}. Since
the doublets play crucial role we called the effect as the {\it doublet blockade}. The effect occurs when the mixing between the doublets $\gamma=0$, which  corresponds to the angle $\theta=\pi/3$ or $\theta=4\pi/3$ (see insert in Fig. \ref{eigenvectors}). The doublet blockade process should be visible in a current characteristic.

To calculate the current we use the diagonalized master equation (DME) which is useful for a finite bias voltage
\cite{poltl}. The equation of motion has the Lindblad form \cite{gurvitz,busl2}
\begin{eqnarray}\label{dme}
\frac{d\rho_{mn}}{dt} &=&-i \langle m|[\widehat{H},\rho]|n\rangle \nonumber \\
 &+& \sum_{k\neq n}(\Gamma_{k\rightarrow n}\rho_{kk}-\Gamma_{n \rightarrow k}\rho_{nn})\delta_{mn} \nonumber \\
 &-& \frac{1}{2}\sum_{k}(\Gamma_{m\rightarrow k}+\Gamma_{n\rightarrow
k})\rho_{mn}(1-\delta_{mn})\,.
\end{eqnarray}
Here the density matrix $\rho$ consists all considered states $m, n \in \{\nu_2, \nu_3\}$, $\rho_{mn}=\langle
m|\rho|n\rangle$, $\Gamma_{n\rightarrow k}=\sum_{\alpha=L,R}(\Gamma^{\alpha+}_{n \rightarrow k}+\Gamma^{\alpha-}_{n \rightarrow k})$ and $\delta_{mn}$ denotes
Kronecker delta.
The first term (\ref{dme}) describes coherent evolution of the qubit in the doublet subspace with the Hamiltonian (\ref{matrix2}), whereas the other terms
correspond to decoherence processes due to coupling with the
electrodes.

The current flowing through the left junction is given by
\begin{eqnarray}\label{curr_general}
I_L=e\sum_{\nu_2,\nu_3}(\Gamma^{L-}_{\nu_3\rightarrow\nu_2}\rho_{\nu_3\nu_3}
-\Gamma^{L+}_{\nu_2\rightarrow\nu_3}\rho_{\nu_2\nu_2}).
\end{eqnarray}
For the stationary case the density matrix elements are derived from the master equation (\ref{dme}) with the left hand side taken as zero. In calculations we
assume that three electron subspace includes doublets as well as quadruplets and for two electrons in TQD the singlet or triplet states are derived from the
Hamiltonian (\ref{HS}) or (\ref{HT}), respectively.

\begin{figure}[ht]
\includegraphics[width=.3\textwidth,clip]{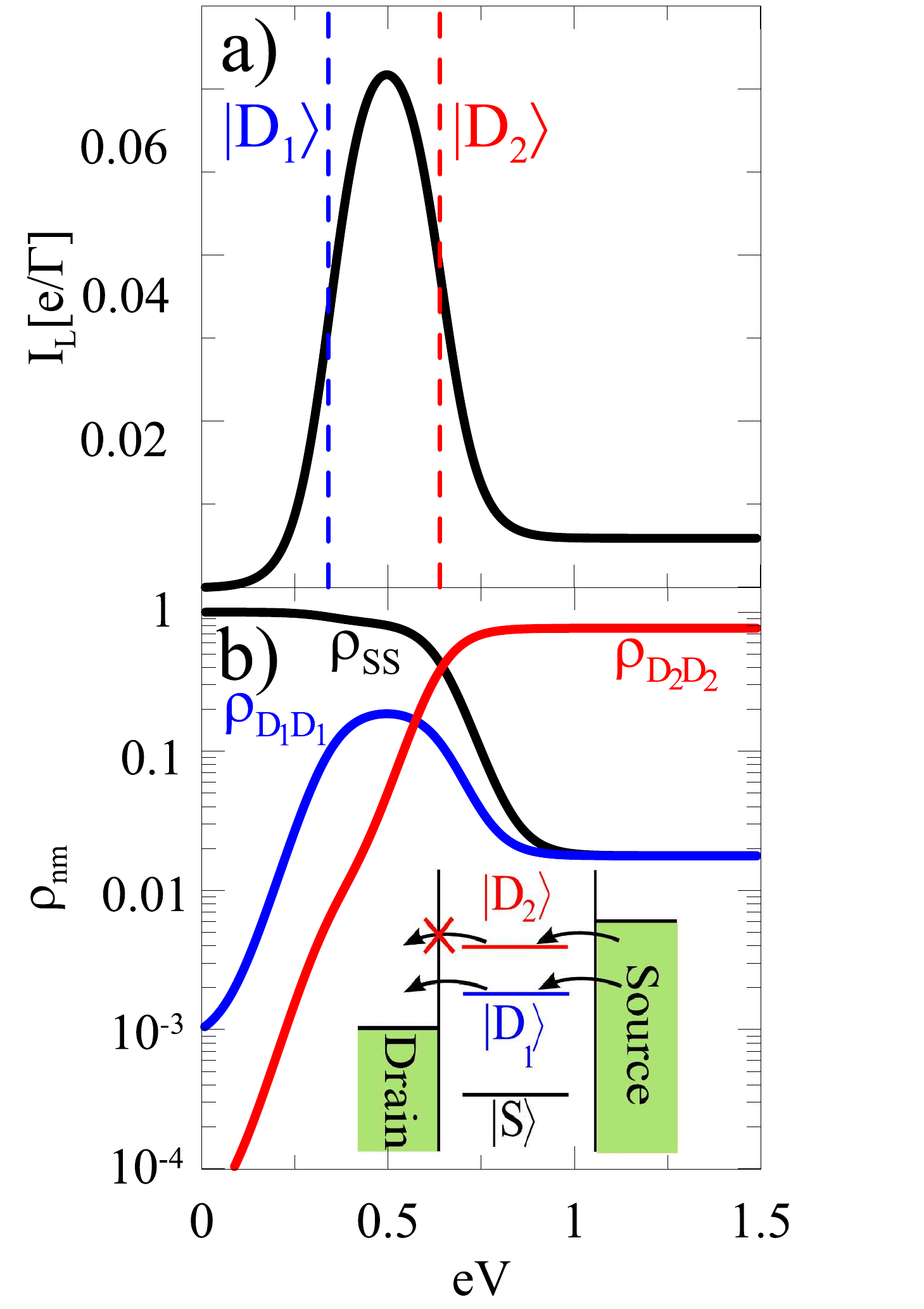}
\caption{Doublet blockade effect for the case with the two-electron ground state as the singlet state.  a) Current characteristic versus applied voltage. The
vertical dashed lines show the
positions of the state \ket{D_1} and \ket{D_2}.  b) Probabilities of occupation of the states: \ket{S} - black, \ket{D_1} - blue and \ket{D_2} - red curve.  The
calculations were performed for the parameters: $J=1$ (in the further calculations it is taken as unity), $\delta=-0.3, \gamma=0.08$ (which corresponds to $g_E=0.2$ and $\theta\approx 1.42 \pi$), $t_{0}=-3, U_{i,i+1}=5$, for which the states are at
$E_S=-1$, $E_{D_1}=13.35$ and $E_{D_2}= 13.65$. The Fermi energy is taken at $E_F=14$, temperature $T=0.05$ and $\Gamma=0.05$.}
\label{blockadeS}
\end{figure}

Numerical calculations were performed for various positions of the Fermi level and the
size of the voltage window. The calculations included all states, but the excited states play a minor role as their
population is thermally activated and is many orders of magnitude smaller. In this paper we confined ourself to transport
studies in the doublet blockade regime and we show how mixing between the doublets \ket{D_{1}} and \ket{D_{2}} removes the current blockade. Fig.\ref{blockadeS}
presents the voltage dependence of the current and the probabilities for occupation of the states \ket{S}, \ket{D_1} and \ket{D_2}.
The Fermi level is set between the states \ket{S} and \ket{D_1} (see the insert in Fig.\ref{blockadeS}b). At the low bias the system is in the Coulomb blockade
regime
and
the current starts to flow at $eV\approx 0.35$ when \ket{D_1} becomes in the voltage window. At a higher voltage $eV\approx 0.65$  one observes a
strong reduction of the current - the doublet blockade effect. This is caused by a high occupation of \ket{D_2} which is uncoupled with the left electrode [see
Eq.(\ref{sd2eq})]. Simultaneously one can see a drop of the occupation of \ket{D_1}. For the case presented in Fig.\ref{blockadeS} we have taken the mixing
parameter
$\gamma=0.08$ in order to show that mixing between  the doublet states removes the current blockade.

\begin{figure}[ht]
\includegraphics[width=.3\textwidth,clip]{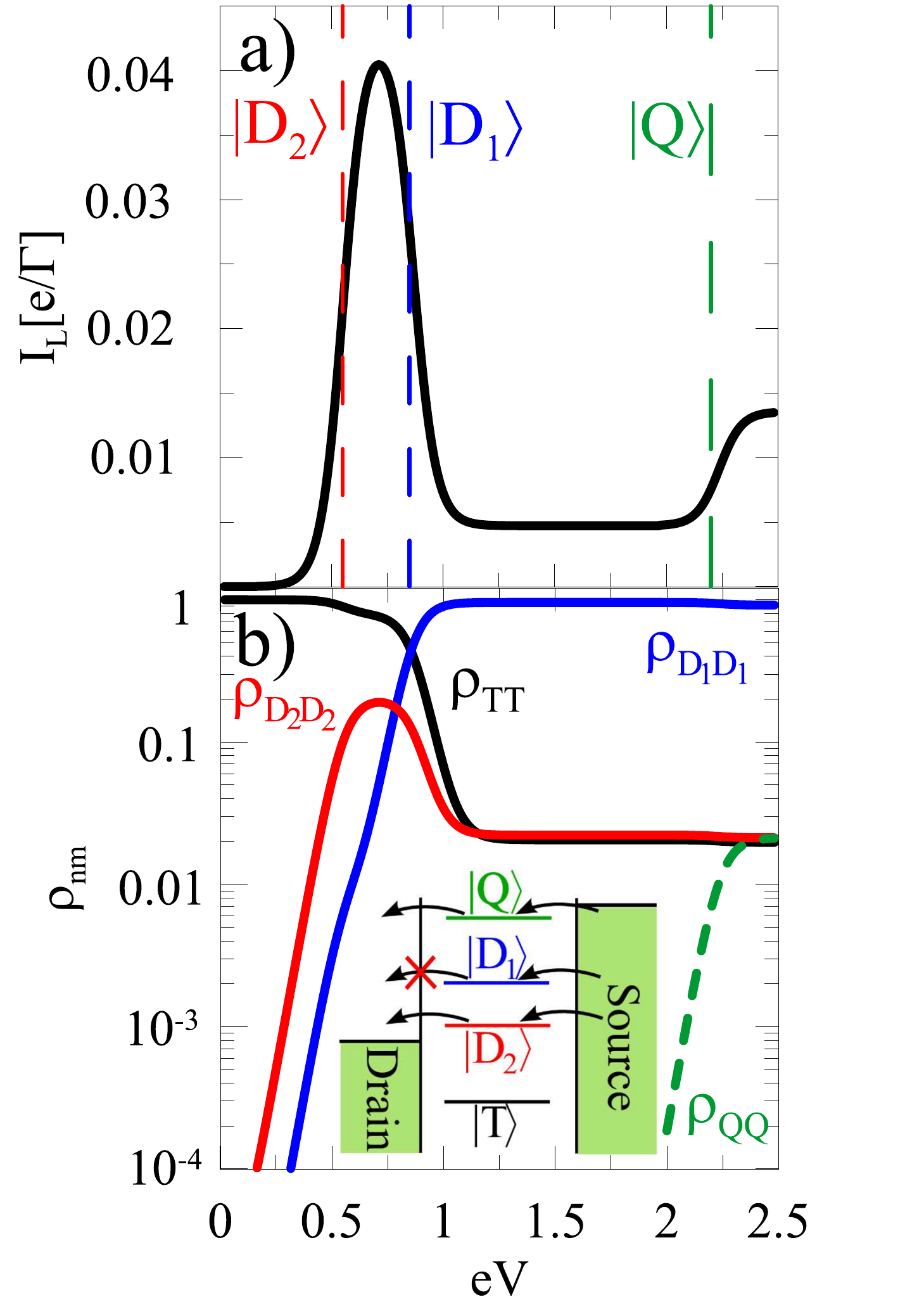}
\caption{Doublet blockade when the two-electron ground state is triplet. a) Current characteristic versus applied voltage. The vertical dashed lines show the
positions of doublets \ket{D_1}, \ket{D_2} as well as quadruplets \ket{Q}.  b) Probabilities of occupation of the states: \ket{T} - black, \ket{D_1} - blue,
\ket{D_2} - red and \ket{Q} - green curve. We use the parameters: $J=1, \delta=0.3, \gamma=0.08, t_{0}=3, U_{ij}=5$, for which
$E_T=-1$, $E_{D_1}=13.65$, $E_{D_2}= 13.35$ and $E_Q=15$. The Fermi energy is $E_F=13.8$, temperature $T=0.05$ and $\Gamma=0.05$.}
\label{blockadeT}
\end{figure}

For comparison we present in Fig.~\ref{blockadeT} the doublet blockade for the case with the triplet as the ground state. Here we have taken the parameter
$\delta$ positive in order to get the transparent state \ket{D_2} to be below the uncoupled state \ket{D_1}. The situation is very similar to the case presented
in Fig.~\ref{blockadeS} but now one can see contribution from quadruplets at higher voltages. In the limit $\gamma\rightarrow 0$ one gets the doublet blockade
regime when all conducting channels are blocked, also those ones through the quadruplet states. If the order of the doublet states is reversed and the uncoupled
state
\ket{D_1} lies below \ket{D_2}, one can observe only a small thermally activated current.

\section{Dynamics - coherent oscillations and relaxation processes} \label{dynamics}

Let us now consider a time evolution of the qubit and its detection by the current measurement. The simplest case is for a moderate magnetic field
which
separates the doublets with different spin orientations. Then one may consider only the evolution in the doublet subspace with the spin $S_z=+1/2$ and ignore
spin-flip processes.

\subsection{Leakage to singlet state}

First we study the case when the current flowing through the system engages the doublet states as well as the singlet state (the ground state for two
electrons).
The dynamic of the system is describe by equation (\ref{dme}), which explicitly has the form:
\begin{eqnarray}\label{master-ds}
\begin{array}{ll}
\dfrac{d}{dt}\rho_{D_1D_1}&=  - \gamma\, \Im\rho_{D_1D_2}-\Gamma^{L-}_{D_1 \to S}\, \rho_{D_1D_1}+\Gamma^{R+}_{S\to D_1}\,\rho_{SS},\\
\dfrac{d}{dt}\rho_{D_2D_2}&=   \gamma\, \Im\rho_{D_1D_2}-\Gamma^{L-}_{D_2 \to S}\, \rho_{D_2D_2}+\Gamma^{R+}_{S\to D_2}\, \rho_{SS},\\
\dfrac{d}{dt}\rho_{D_1D_2}&=  -i \delta\,\rho_{D_1D_2}-i \dfrac{\gamma}{2}\,
(\rho_{D_2D_2}-\rho_{D_1D_1})\\&-\frac{1}{2}(\Gamma^{L-}_{D_1 \to S}+\Gamma^{L-}_{D_2 \to S})\,\rho_{D_1D_2},\\
\dfrac{d}{dt}\rho_{SS}&= \Gamma^{L-}_{D_1\to S}\,\rho_{D_1D_1}+\Gamma^{L-}_{D_2\to S}\,\rho_{D_2D_2}\\&-(\Gamma^{R+}_{S\to D_1}+\Gamma^{R+}_{S\to
D_2})\,\rho_{SS}\,.
      \end{array}
\end{eqnarray}
To simplify the notation the spin index in the doublet states is omitted.
We assume that the both doublet states are in the voltage window. For low temperatures one can take into account only electron transfers from the right to the
left
hand side and ignore back transfers. As we noted above the tunneling rates on the left and the right junction enter the master equation in an asymmetric way.
For
our case $\Gamma^{L-}_{D_2\to S}=0$ and $\Gamma^{L-}_{D_1\to S}\neq0$ which describe the decay of the resonant states, whereas $\Gamma^{R+}_{S \to D_1}$ and
$\Gamma^{R+}_{S \to D_2}$ describe the build-up of these states. Since the system is in the doublet blockade regime an electron can be pumped to the state
\ket{D_2} but it can not leave this state in the absence of the mixing term (for $\gamma=0$).
These equations are similar to those ones for single-spin dynamics in a quantum dot in the case of the Pauli spin blockade \cite{engel} (see also \cite{stoof}).

\begin{figure}
\includegraphics[width=.3\textwidth,clip]{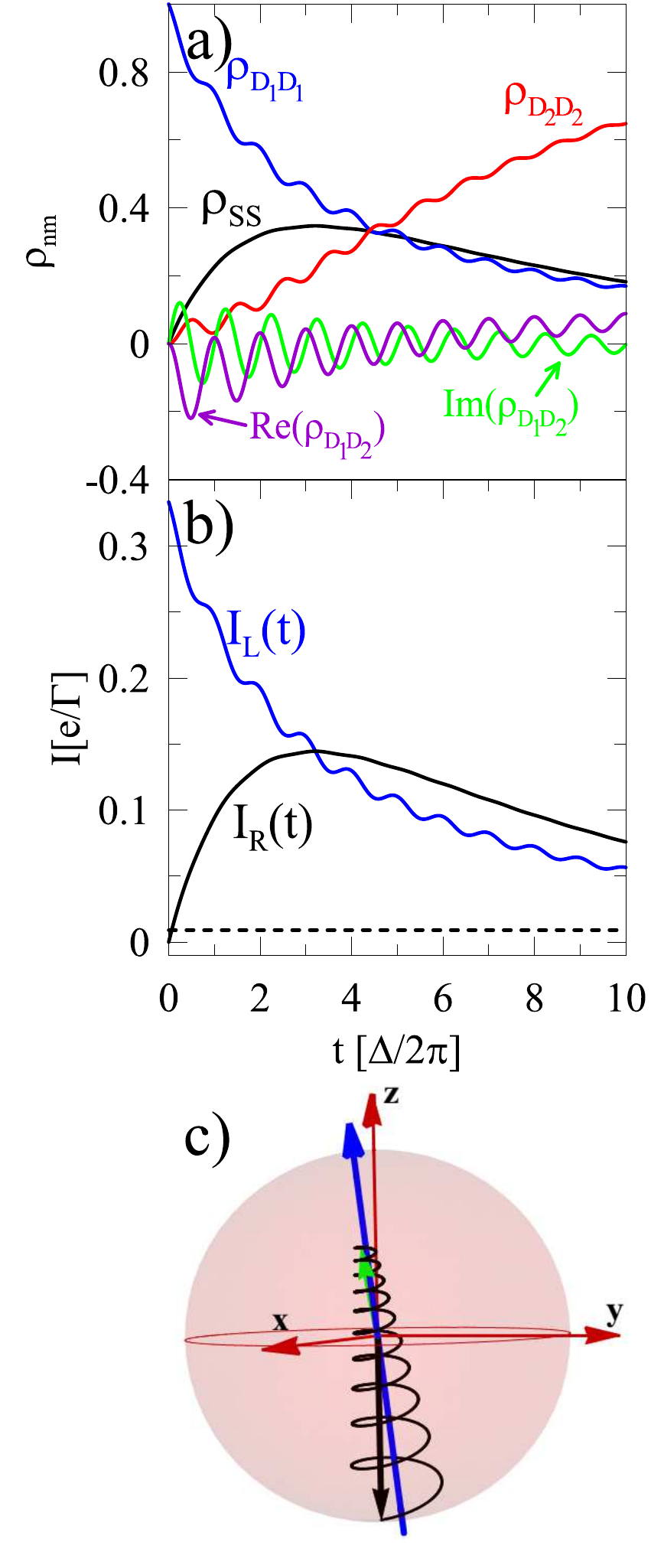}
\caption{Coherent oscillations when the singlet ground state is engaged in the current flow. The plots are derived for the time independent mixing
$\gamma(t)=\gamma_0$ of the doublet states \ket{D_1} and \ket{D_2}. Panel a) and b) present: the density matrix elements and the currents flowing through the
junctions, respectively. The black-dashed line corresponds to the current in the stationary limit. Panel c) shows the time evolution of the pseudo-spin on the
Bloch sphere. The black arrow represents the initial state \ket{D_1} and the green arrow - the final state. The parameters taken in the calculations:
$\delta=-0.3$,
$\gamma_0=0.08, \Gamma=0.05, \rho_{D_1D_1}(0)=1$, and the other parameters as in Fig.\ref{blockadeS}. The relaxation rates calculated exact are: $1/T_{leak}=0.0306$, $1/T_1=0.0069$, and $1/T_2=0.0083$.}
\label{singdynamic}
\end{figure}

The qubit is prepared either in the state \ket{D_1} or \ket{D_2} as described in the chapter \ref{preparation}. Next at the initial time $t=0$ the mixing term
becomes switched on (by changing the orientation of the effective electric field). We consider first the case for the time independent mixing term,
$\gamma=\gamma_0$ for $t\geq 0$.

In order to have better insight into relaxation processes the equations (\ref{master-ds}) are rewritten in the Bloch vector base:
\begin{align}\label{bloch-sing}
\begin{array}{ll}
\dfrac{d}{dt}s_x&=-\delta s_y-\frac{1}{2}\Gamma^{L-}_{D_1 \to S}\,s_x ,\\
\dfrac{d}{dt}s_y&=\delta s_x+\gamma\, s_z-\frac{1}{2}\Gamma^{L-}_{D_1 \to S}\,s_y ,\\
\dfrac{d}{dt} s_z&=-\gamma\,s_y-\frac{1}{2}\Gamma^{L-}_{D_1 \to S}\,s_{z}\\&-\left(\Gamma^{R+}_{S \to D_1}-\Gamma^{R+}_{S \to D_2}+\frac{1}{2}\Gamma^{L-}_{D_1
\to
S}\right)\,\rho_{SS}+\frac{1}{2}\Gamma^{L-}_{D_1 \to S} ,\\
\dfrac{d}{dt}\rho_{SS}&= -\frac{1}{2}\Gamma^{L-}_{D_1 \to S}\,s_z \\&- \left(\Gamma^{R+}_{S \to D_1} +\Gamma^{R+}_{S \to D_2}+\frac{1}{2}\Gamma^{L-}_{D_1 \to S}
\right)\,\rho_{SS}+\frac{1}{2}\Gamma^{L-}_{D_1 \to S}.
\end{array}
\end{align}
Here, the vector components are $s_x=\rho_{D_1D_2}+\rho_{D_2D_1}$, $s_y=i(\rho_{D_1D_2}-\rho_{D_2D_1})$  and $s_z=\rho_{D_2D_2}-\rho_{D_1D_1}$.
We also used the condition $\rho_{D_1D_1}(t)+\rho_{D_2D_2}(t)+\rho_{SS}(t)=1$, which is fulfilled for any time. These equations are similar to the optical Bloch
equations and therefore, by analogy, we may take [from the first two equations in (\ref{bloch-sing})] a decoherence rate $1/T_2\approx \Gamma^{L-}_{D_1 \to S}/2$. This rate describes how fast the superposition of
states \ket{D_1} and \ket{D_2} loss the coherence due to interactions with the electrodes. From the third equation in (\ref{bloch-sing}) one can find a relaxation rate  $1/T_1
\approx  \Gamma^{L-}_{D_1 \to S}/2$ which describes evolution of z-component of the pseudo-spin. In contrast for a standard optical Bloch
equations one expects $1/T_1=2/T_2$. In our case $1/T_1=1/T_2$, because they are caused only by coupling with electrodes, and we neglect thermalization processes
in the quantum dot system.

There is also an additional term describing a leakage from the qubit space to the \ket{S} state, which causes the collapse of the Bloch sphere.
The relaxation rate for the leakage process is $1/T_{leak}\approx \left(\Gamma^{R+}_{S \to D_1}+\Gamma^{R+}_{S \to D_2}+\Gamma^{L-}_{D_1 \to S}/2\right)$ [see the
fourth equation in (\ref{bloch-sing})].

The relaxation rates $1/T_1$, $1/T_2$ and $1/T_{leak}$ considered above contain the main contribution parts only. To get full information about the rates one
needs to solve exactly the
differential equations (\ref{master-ds}) or (\ref{bloch-sing}). We solved eq. (\ref{master-ds})  by means of the Laplace transformation $\hat{g}(z)=\int_0^\infty
\hat{\rho}(t) \,\exp(-z t)
dt$,
where $\hat{\rho}=\{\rho_{D_1D_1}, \rho_{D_2D_2}, \rho_{SS}, \Re\rho_{D_1D_2}, \Im\rho_{D_1D_2}\}$. This method is very useful, because the poles of
$\hat{g}(z)$
give information on the relaxation rates (real parts of the poles) and on frequencies of eigenmodes (imaginary parts of the poles).
Although one can get analytical solutions in our case, they are too complex and illegible, therefore we present numerical results only.
In the final step we made the inverse Laplace transformation to get the time evolution of the system $\hat{\rho}(t)$. From  (\ref{curr_general}) we get currents
flowing through the left and the right tunnel junction:
\begin{eqnarray}\label{currL}
I_L(t)=e\Gamma^{L-}_{D_1\to S}\,\rho_{D_1D_1}(t),\\ \label{currR}
 I_R(t)=e(\Gamma^{R+}_{S\to D_1}+\Gamma^{R+}_{S\to D_2})\rho_{SS}(t).
\end{eqnarray}
Notice that in general $I_L(t)\neq I_R(t)$, which exhibits time dependent charge accumulation in the system. Of course in the stationary limit ($t\to \infty$)
\begin{widetext}
\begin{eqnarray}
I^0_L=I^0_R=\frac{e\gamma_0^2\,\Gamma^{L-}_{D_1\to S}(\Gamma^{R+}_{S\to D_1} +\Gamma^{R+}_{S\to D_2})}{\gamma_0^2 (2
\Gamma^{R+}_{S\to D_1} + 2 \Gamma^{R+}_{S\to D_2} +\Gamma^{L-}_{D_1\to S}) + \Gamma^{R+}_{S\to D_2} (4 \delta^2 +
(\Gamma^{L-}_{D_1\to S})^2)}.
\label{statlimit}
\end{eqnarray}
\end{widetext}

The numerical results for $\rho_{ij}(t)$ are presented in Fig. \ref{singdynamic}a. We have assumed that the qubit is
prepared in the state \ket{D_1}. One can see that the population of the singlet state ($\rho_{SS}$) increases at
the beginning of the measurement. It is the effect of leakage from the doublet subspace to the singlet state with the relaxation time $T_{leak}$.
For longer times $\rho_{SS}$ decrease with the relaxation time $T_1$. Because the current $I_R(t)$ is proportional to $\rho_{SS}$, eq. (\ref{currR}), these processes can be measured
in the short and long time range, respectively - see fig.\ref{singdynamic}b.
One can see also (fig. \ref{singdynamic}a) that the occupation of the state \ket{D_1} decreases, whereas \ket{D_2} increases. It is related with trapping of an
electron in the dark state \ket{D_2} (the doublet blockade effect).
The quantities $\rho_{D_1D_1}(t)$ and
$\rho_{D_2D_2}(t)$ reach their stationary values with the relaxation rate $1/T_1$. The diminish of \ket{D_1} can be directly seen in the $I_L(t)$
characteristic  (fig.\ref{singdynamic}b). One can say that measurement of the current flowing through the left and the right junction gives information about
dynamics and relaxation processes in the qubit subspace.

The oscillations of $\rho_{D_1D_1}$ and $\rho_{D_2D_2}$ are related with the coherent oscillations which can be seen for the curves $\Re{\rho_{D1D2}}$ and
$\Im{\rho_{D1D2}}$. The period of the oscillations is equal to $\Delta/(2\pi)$,
whereas their amplitude is $[\gamma_0/(2 \Delta)]^2$ and decreases with the decoherence rate $1/T_2$. In fig. \ref{singdynamic}c we present these oscillations as
a rotation of the pseudo-spin vector on the Bloch sphere. For the initial conditions the pseudo-spin
points out the south pole (a black arrow). The final state is represented as a green arrow and it deviated from the z-axis, because of nonzero mixing $\gamma_0$. One can see the pseudo-spin rotates
on the helix, which radius is diminished in time due to decoherence with the time $T_2$ and it is called as a phase damping. The axis of rotation (blue arrow) is
given by $\left(-\delta \gamma_0/\Delta^2,0,\delta^2/\Delta^2\right)$.
Another damping is related with relaxation to the stationary state with the rate $1/T_1$ -- this is called as the amplitude damping \cite{nielsen}. The leakage
process reveals itself in the short time scale (in two first cycles) and the effect is clearly seen in the measurement of the current $I_R(t)$.

Now we analyze the driven case for an AC electric field which cause oscillation of mixing term $\gamma(t)=\gamma_0\exp(-i \omega t)$.
Dynamics is described by the equations of motion Eq.(\ref{dme}) in the rotating frame (RF) approximation, in which $|D_1\rangle_{RF}=|D_1\rangle \exp(i\omega
t/2)$ and $|D_2\rangle_{RF}=|D_2\rangle \exp(-i\omega t/2)$.  The stationary current is given by
\begin{widetext}
\begin{eqnarray}
I^0_L(\omega)=I^0_R(\omega)=\frac{e\gamma_0^2\,\Gamma^{L-}_{D_1\to S}(\Gamma^{R+}_{S\to D_1} +\Gamma^{R+}_{S\to D_2})}{\gamma_0^2 (2
\Gamma^{R+}_{S\to D_1} + 2 \Gamma^{R+}_{S\to D_2} +\Gamma^{L-}_{D_1\to S}) + \Gamma^{R+}_{S\to D_2} [4 (\delta-\omega)^2 +
(\Gamma^{L-}_{D_1\to S})^2]}.
\label{statlimitrez}
\end{eqnarray}
\end{widetext}
$I^0_L(\omega)$ has a Lorentzian dependence and reaches its maximum for a resonance condition $\omega=\delta$. In this case the doublet blockade is
removed and higher current flows through the system.

\begin{figure}
\includegraphics[width=.3\textwidth,clip]{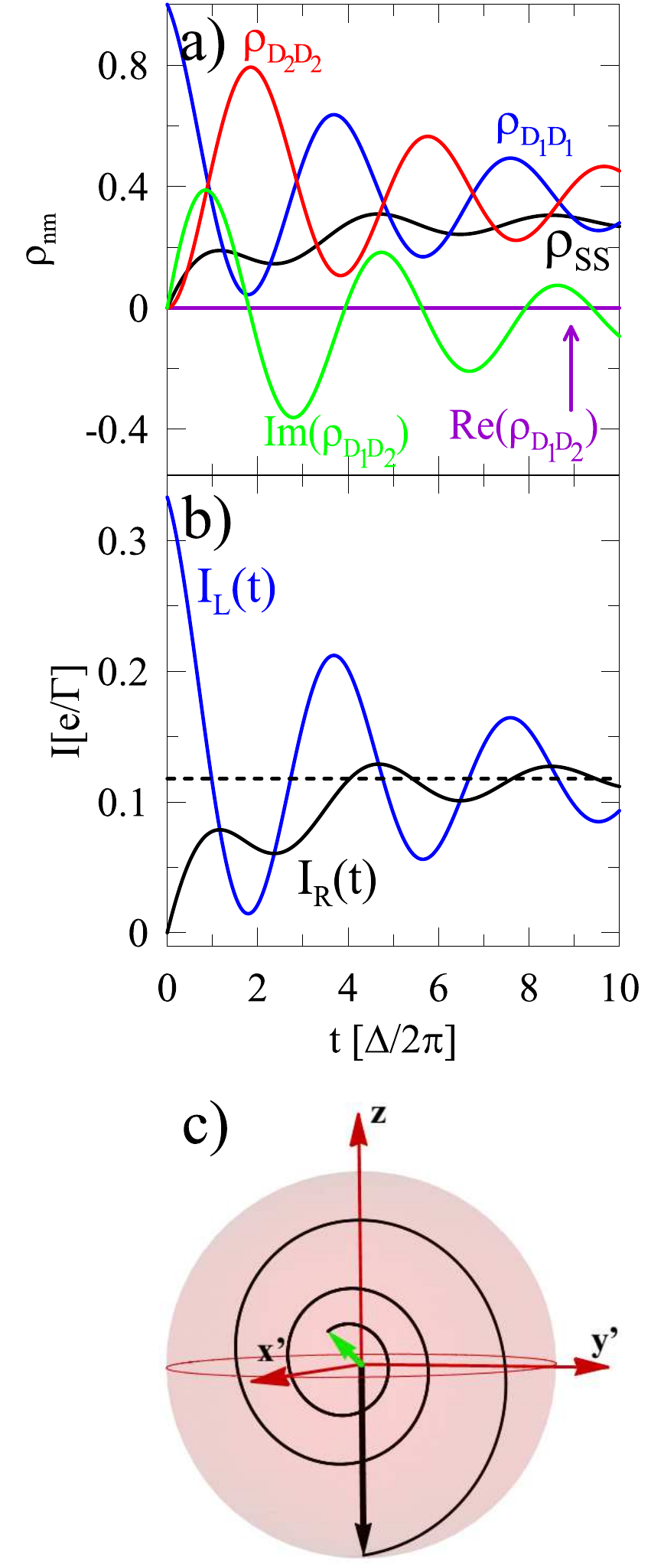}
\caption{Rabi oscillations for the driven case $\gamma(t)=\gamma_0\exp(-i \omega t)$ at the resonance condition ($\omega=\delta$). The notation and
the parameters are the same as in
Fig.~\ref{singdynamic}. Notice that the results are presented in the rotating frame, whereas Fig.~\ref{singdynamic} presents the
evolution in the laboratory frame. The relaxation rates are $1/T_{leak}=0.0306$, and $1/T_2=0.0083$.}
\label{singdynamicrez}
\end{figure}

Fig.~\ref{singdynamicrez} shows the density matrix elements, the currents and the oscillations on the Bloch sphere for the driven case in the resonance ($\omega=\delta$). We take the same parameters as for Fig.~\ref{singdynamic} but now the results are presented in the rotating frame.
One can see the large Rabi oscillations of $\rho_{D_1D_1}$ and $\rho_{D_2D_2}$.
The population of the states is changed alternately between \ket{D_1} and \ket{D_2} with the frequency $|\gamma_0|$. It is also seen in Fig.
\ref{singdynamicrez}c which presents the rotation of the pseudo-spin on the Bloch sphere. The rotation is in the $y'$-$z$ plane on the spiral with periodic
transfers between the states $\rho_{D_1D_1}$ and $\rho_{D_2D_2}$. In the stationary limit one gets $\rho_{D_1D_1}(\infty)=\rho_{D_2D_2}(\infty)=0.42$ and
$\rho_{SS}(\infty)=0.16$, which is a higher value than in the non-driven case with $\rho_{SS}(\infty)=0.02$ (see Fig. \ref{singdynamic}). The relaxation times
$1/T_2$ and $1/T_{leak}$ are almost the same as for the time independent case due to their weak dependance on $\gamma$ and $\delta$.

The current plots in Fig.~\ref{singdynamicrez}b) present strong oscillations which corresponds to coherent switching between the doublet states (Rabi
oscillations).
The leakage current flowing through the right junction $I_R$ also shows some oscillations. The stationary current (dashed line) is larger than for the
non-driven
case as one may expect when the doublet blockade is removed.

\subsection{Leakage to triplet and quadruplet states}

\begin{figure}
\includegraphics[width=.3\textwidth,clip]{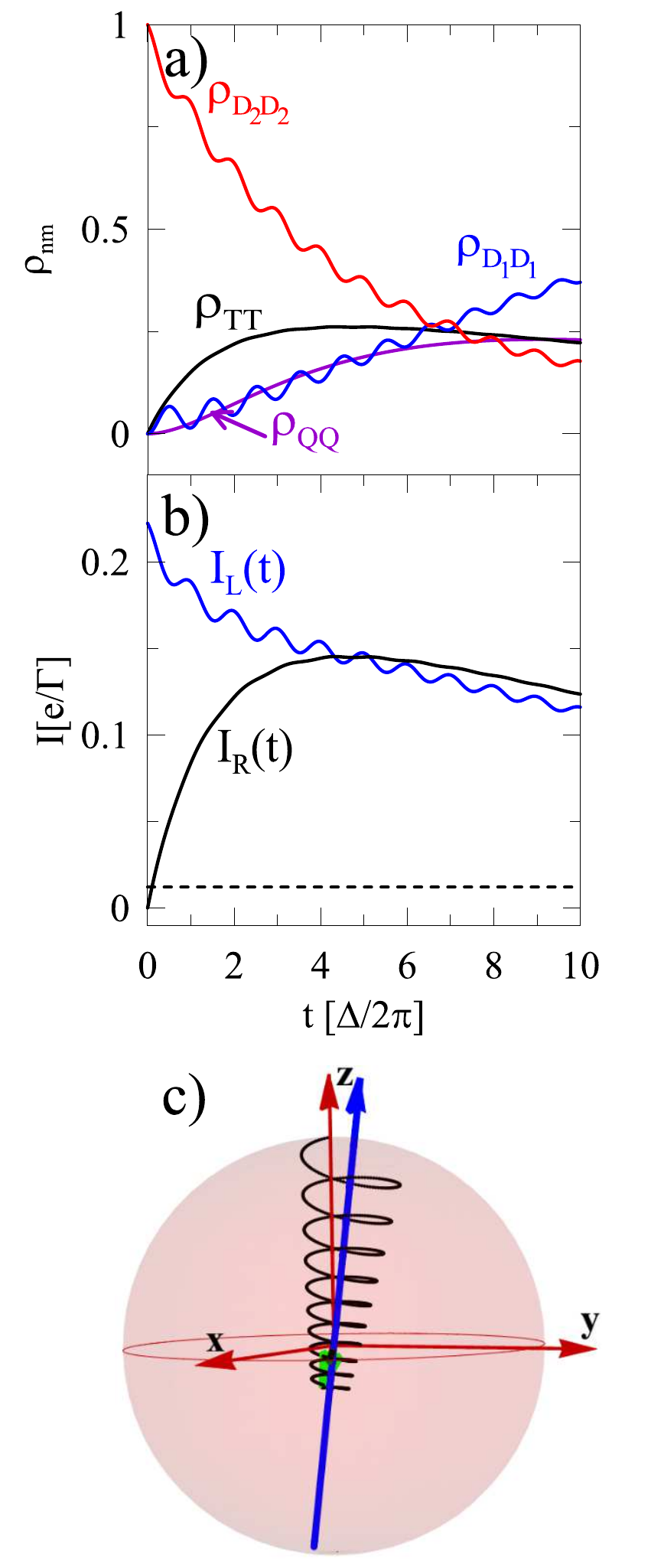}
\caption{Coherent oscillations when the triplet ground state is engaged in the current flow. Figures are plotted  for the time independent mixing term with
$\gamma(t)=\gamma_0$ switched on at $t=0$. Panel a) presents the time evolution of the density matrix elements, while b) presents the current flowing through
the
left (blue curve) and right junction (black curve). The black-dashed line corresponds
to the current in the stationary limit. Panel c) shows the evolution of the pseudo-spin on the Bloch sphere. The initial state is \ket{D_2} (the black
arrow), and the final state is represented by the green arrow. The parameters taken in the calculations: $\delta=0.3$, $\gamma_0=0.08, \Gamma=0.05.$ The other
parameters are the same as in Fig.\ref{blockadeT}. The relaxation rates for these parameters are $1/T_{leak}=0.0405$, $1/T^{TQ}_{leak}=0.0117$,
$1/T_1=0.0033 $, and $1/T_2=0.0055$. Notice that in the present case the currents are smaller than those ones in Fig.\ref{singdynamic}b because now the transfer
rates are different.}
\label{trypdynamic}
\end{figure}

One can expect similar dynamics when the two-electron ground state is triplet. Here we still confine ourselves to the doublets with $S_z=1/2$ and  ignore
spin-flip processes. Now in the voltage window (see the inset in Fig.\ref{trypdynamic}) we have the states
\ket{D^{1/2}_1}, \ket{D^{1/2}_2} and \ket{Q^{3/2}} whereas \ket{T^1} is below the chemical potential in the left electrode. The Master equation (\ref{dme}) is
rewritten in the form
\begin{align}\label{master-dT}
\begin{array}{ll}
\dfrac{d}{dt}\rho_{D_1D_1}&=  -  \gamma\, \Im\rho_{D_1D_2}+\Gamma^{R+}_{T\to D_1}\, \rho_{TT},\\
\dfrac{d}{dt}\rho_{D_2D_2}&=   \gamma\, \Im\rho_{D_1D_2}-\Gamma^{L-}_{D_2 \to T}\, \rho_{D_2D_2}+\Gamma^{R+}_{T\to
D_2}\,\rho_{TT},\\
\dfrac{d}{dt}\rho_{D_1D_2}&=  -i \delta\,\rho_{D_1D_2}-i \dfrac{\gamma}{2}\,
(\rho_{D_2D_2}-\rho_{D_1D_1})\\&-\frac{1}{2}\Gamma^{L-}_{D_2 \to T}\,\rho_{D_1D_2},\\
\dfrac{d}{dt}\rho_{TT}&= \Gamma^{L-}_{D_2\to T}\,\rho_{D_2D_2} +\Gamma^{L-}_{Q\to T}\,\rho_{QQ}\\&-\left(\Gamma^{R+}_{T\to D_1}+\Gamma^{R+}_{T\to
D_2}+\Gamma^{R+}_{T\to Q}\right)\,\rho_{TT},\\
\dfrac{d}{dt}\rho_{QQ}&= \Gamma^{R+}_{T\to Q}\,\rho_{TT} -   \Gamma^{L-}_{Q\to T}\,\rho_{QQ}.
      \end{array}
\end{align}
Here the population of the states is built-up by the transfers $\Gamma^{R+}_{T\to D_1}$, $\Gamma^{R+}_{T\to D_2}$ and $\Gamma^{R+}_{T\to Q}$ from the right
electrode.
Electrons escape from the system to the left electrode which is described by $\Gamma^{L-}_{D_2\to T}$ and $\Gamma^{L-}_{Q\to T}$. The currents flowing from the
right and the left electrode are expressed as
\begin{eqnarray}\label{currL_T}
I_L(t)=e(\Gamma^{L-}_{D_2 \to T}\,\rho_{D_2D_2}+\Gamma^{L-}_{Q \to
T}\,\rho_{QQ})\\ \label{currR_T}
I_R(t)=e
(\Gamma^{R+}_{T \to D_1} + \Gamma^{R+}_{T \to D_2}+ \Gamma^{R+}_{T \to Q})\,\rho_{TT}.
\end{eqnarray}
Notice that now the role of the doublet states \ket{D_1} and \ket{D_2} is reversed, and current flows through \ket{D_2} while \ket{D_1} is the dark state and
blocks electron transport - see Eq.(\ref{DtoT1}).
In the Bloch space we have:
\begin{align}\label{bloch-trip}
\begin{array}{ll}
\dfrac{d}{dt}s_x&=-\delta s_y-\frac{1}{2}\Gamma^{L-}_{D_2 \to T}\,s_x \\
\dfrac{d}{dt}s_y&=\delta s_x+\gamma\, s_z-\frac{1}{2}\Gamma^{L-}_{D_2 \to T}\,s_y \\
\dfrac{d}{dt} s_z&=-\gamma\,s_y-\frac{1}{2}\Gamma^{L-}_{D_2 \to T}\,s_{z} \\ & -\left(\Gamma^{R+}_{T \to D_1}-\Gamma^{R+}_{T \to D_2}-\frac{1}{2}\Gamma^{L-}_{D_2
\to
T}\right)\,\rho_{TT} \vspace{0.1cm} \\ & +\frac{1}{2}\Gamma^{L-}_{D_2 \to T}\,\rho_{QQ}-\frac{1}{2}\Gamma^{L-}_{D_2 \to T}\\
\dfrac{d}{dt}\rho_{TT}&= \frac{1}{2}\Gamma^{L-}_{D_2 \to T}\,s_z \\ & - \left(\Gamma^{R+}_{T \to D_1} +\Gamma^{R+}_{T \to D_2}+\Gamma^{R+}_{T \to
Q}+\frac{1}{2}\Gamma^{L-}_{D_2 \to T} \right)\,\rho_{TT} \vspace{0.1cm} \\ & +\left(\Gamma^{L-}_{Q \to T}-\frac{1}{2}\Gamma^{L-}_{D_2 \to
T}\right)\,\rho_{QQ}+\frac{1}{2}\Gamma^{L-}_{D_2 \to T}\\
\dfrac{d}{dt}\rho_{QQ}&=\Gamma^{R+}_{T \to Q}\,\rho_{TT}-\Gamma^{L-}_{Q \to T}\,\rho_{QQ}
\end{array}
\end{align}
One can easily find the main contributions to the relaxation rates:
$1/T_1\approx 1/T_2 \approx \frac{1}{2}\Gamma^{L-}_{D_2 \to T}$, which is qualitatively similar to the previously considered case with the singlet state but now
the transfer rates are different [compare Eq.(\ref{DtoS})-(\ref{DtoS2}) with Eq.(\ref{DtoT1})-(\ref{DtoT4})].
The leakage to the triplet state is given by $1/T_{leak}\approx \Gamma^{R+}_{T \to D_1}
+\Gamma^{R+}_{T \to D_2}+\Gamma^{R+}_{T \to Q}+\frac{1}{2}\Gamma^{L-}_{D_2 \to T}$ which has an additional term $\Gamma^{R+}_{T \to Q}$ describing transfer from
the triplet to quadruplet state.
The last row in Eq.(\ref{master-dT}) and (\ref{bloch-trip}) describes another leakage process with the relaxation rate $1/T^{TQ}_{leak}\approx \Gamma^{L-}_{Q \to
T}$.
This process changes the population of the triplet and the quadruplet state  and indirectly influences dynamics in the doublet subspace.

Solving these equations we determine the occupation probabilities and consequently the currents: $I_L(t)$ and $I_R(t)$. The results
are presented in Fig.~\ref{trypdynamic} for the time independent mixing term, $\gamma(t)=\gamma_0$ for $t\geq 0$ and \ket{D_2} as the initial state.
On the top panel one can see coherent oscillations for $\rho_{D_1D_1}$ and $\rho_{D_2D_2}$  similar as for the case with the singlet state.
However, due to the quadruplet state the dynamic of the doublets and their final occupation is different than for the singlet case. In the short time range the
population $\rho_{QQ}$ and $\rho_{TT}$ increases due to leakage from the doublet subspace with characteristics rates $1/T_{leak}$ and $1/T^{TQ}_{leak}$.

In the longer time scale the population $\rho_{QQ}$ goes to $\rho_{TT}$ (in the stationary limit $\rho_{QQ}(\infty)\approx\rho_{TT}(\infty)=0.021$).
The influence of these states on the doublet dynamics is clearly seen in the currents $I_L$ and $I_R$ presented in Fig. \ref{trypdynamic}b.
$I_L(t)$ shows different behavior than in the singlet case, because the doublet blockade is modified by the quadruplet state which gives an additional
contribution to the current. The increase of the current $I_R(t)$ at the short time scale is related with the leakage to the triplet state.
For longer times $I_R(t)$ is diminished but the quadruplet contribution makes
the drop less pronounced than in the singlet case.

The bottom panel, Fig. \ref{trypdynamic}c, presents the doublet dynamics on the Bloch sphere. The behavior of the Bloch vector is similar as in the singlet case
with some quantitative differences. The phase and the amplitude damping is smaller which is the result of the longer relaxation times in the considered case.

\subsection{Spin-flip processes}

In the considerations above we have taken into account only charge fluctuations on the evolution of the qubit. Let us now extend the studies and include spin-flip
processes. A spin of an electron captured on a quantum dot can interact with nuclear spins of many atoms confined in the area of the quantum dot, which can lead
to decoherence of the qubit states. The decoherence processes due to hyperfine interaction in triangular spin clusters has been already investigated by Troiani et al. \cite{troiani}.
Here we consider another decoherence process caused by the spin relaxation in the electrodes. The electrodes connected to TQD are paramagnetic and electron can be injected with spin up to the state \ket{D^{1/2}_{1,2}} or with spin down to \ket{D^{-1/2}_{1,2}}. This stochastic process leads to mixing between two doublet subspaces.

The evolution of the qubit is studied in the absence of the magnetic field. We assume that the qubit is prepared in the state
\ket{D^{1/2}_1} with the spin $S_z=+1/2$ and the singlet is the ground state for two electrons.
Similarly as in in the previous cases the qubit dynamics is govern by the Master equation (\ref{dme}), but now we take into account states with different spin
orientation \ket{D^{\pm 1/2}_1}, \ket{D^{\pm 1/2}_2}  and \ket{S}
\begin{widetext}
\begin{eqnarray}\label{master-spin-flip}
\begin{array}{ll}
\dfrac{d}{dt}\rho_{D_1^{1/2}D_1^{1/2}}&=  -\gamma \Im \rho_{D_1^{1/2}D_2^{1/2}}-\Gamma^{L-}_{D_1 \to S}\,\rho_{D_1^{1/2}D_1^{1/2}}+\Gamma^{R+}_{S\to
D_1}\,\rho_{SS}\\
\dfrac{d}{dt}\rho_{D_2^{1/2}D_2^{1/2}}&=  \gamma \Im \rho_{D_1^{1/2}D_2^{1/2}}+\Gamma^{R+}_{S\to
D_2}\,\rho_{SS}\\
\dfrac{d}{dt}\rho_{D_1^{1/2}D_2^{1/2}}&=  - i \delta\,\rho_{D_1^{1/2}D_2^{1/2}}-i
\frac{\gamma}{2}\left(\rho_{D_2^{1/2}D_2^{1/2}}-\rho_{D_1^{1/2}D_1^{1/2}}\right)-  \frac{1}{2} \Gamma^{L-}_{D_1 \to
S}\,\rho_{D_1^{1/2}D_2^{1/2}}\\
\dfrac{d}{d}\rho_{SS}&=\Gamma^{L-}_{D_1 \to S}\left(\rho_{D_1^{1/2}D_1^{1/2}}+\rho_{D_1^{-1/2}D_1^{-1/2}}\right)-2\left(\Gamma^{R+}_{S \to D_1}+\Gamma^{R+}_{S
\to D_2}\right)\rho_{SS}\\
\dfrac{d}{dt}\rho_{D_1^{-1/2}D_1^{-1/2}}&=  -\gamma \Im \rho_{D_1^{-1/2}D_2^{-1/2}}-\Gamma^{L-}_{D_1 \to S}\,\rho_{D_1^{-1/2}D_1^{-1/2}}+\Gamma^{R+}_{S\to
D_1}\,\rho_{SS}\\
\dfrac{d}{dt}\rho_{D_2^{-1/2}D_2^{-1/2}}&=  \gamma \Im \rho_{D_1^{-1/2}D_2^{-1/2}}+\Gamma^{R+}_{S\to
D_2}\,\rho_{SS}\\
\dfrac{d}{dt}\rho_{D_1^{-1/2}D_2^{-1/2}}&=  - i \delta\,\rho_{D_1^{-1/2}D_2^{-1/2}}-i
\frac{\gamma}{2}\left(\rho_{D_2^{-1/2}D_2^{-1/2}}-\rho_{D_1^{-1/2}D_1^{-1/2}}\right)-  \frac{1}{2} \Gamma^{L-}_{D_1 \to
S}\,\rho_{D_1^{-1/2}D_2^{-1/2}}\\
\end{array}
\end{eqnarray}
\end{widetext}
where the transfer rates are the same for both spin orientations: $\Gamma^{L-}_{D_n \to S}=\Gamma^{L-}_{D^{1/2}_n \to S}=\Gamma^{L-}_{D^{-1/2}_n \to S}$ and $\Gamma^{R+}_{S \to D_n}=\Gamma^{R+}_{S \to
D^{1/2}_n }=\Gamma^{R+}_{S \to D^{-1/2}_n}$. One can see that the Master equation (\ref{master-spin-flip}) represents dynamics of two doublet subspaces with
$S_z=1/2$ and $S_z=-1/2$. The subspaces are mixed with each other by transfers to the singlet state [see fourth equation in (\ref{master-spin-flip})] . These two subspaces correspond to two pseudo-spin vectors on two Bloch spheres. Each of the
subspace is described by the Master equation Eq. (\ref{bloch-sing}) but the relaxation rates are now: $1/T_1\approx1/T_2\approx\Gamma^{L-}_{D_1 \to S}/2$ and the
leakage process $1/T_{leak}\approx 2\Gamma^{R+}_{S \to D_1}+2\Gamma^{R+}_{S \to D_2}+\Gamma^{L-}_{D_1 \to S}$, which is twice larger than in the case without
spin-flip due states degeneracy.

\begin{figure}[ht]
\includegraphics[width=.3\textwidth,clip]{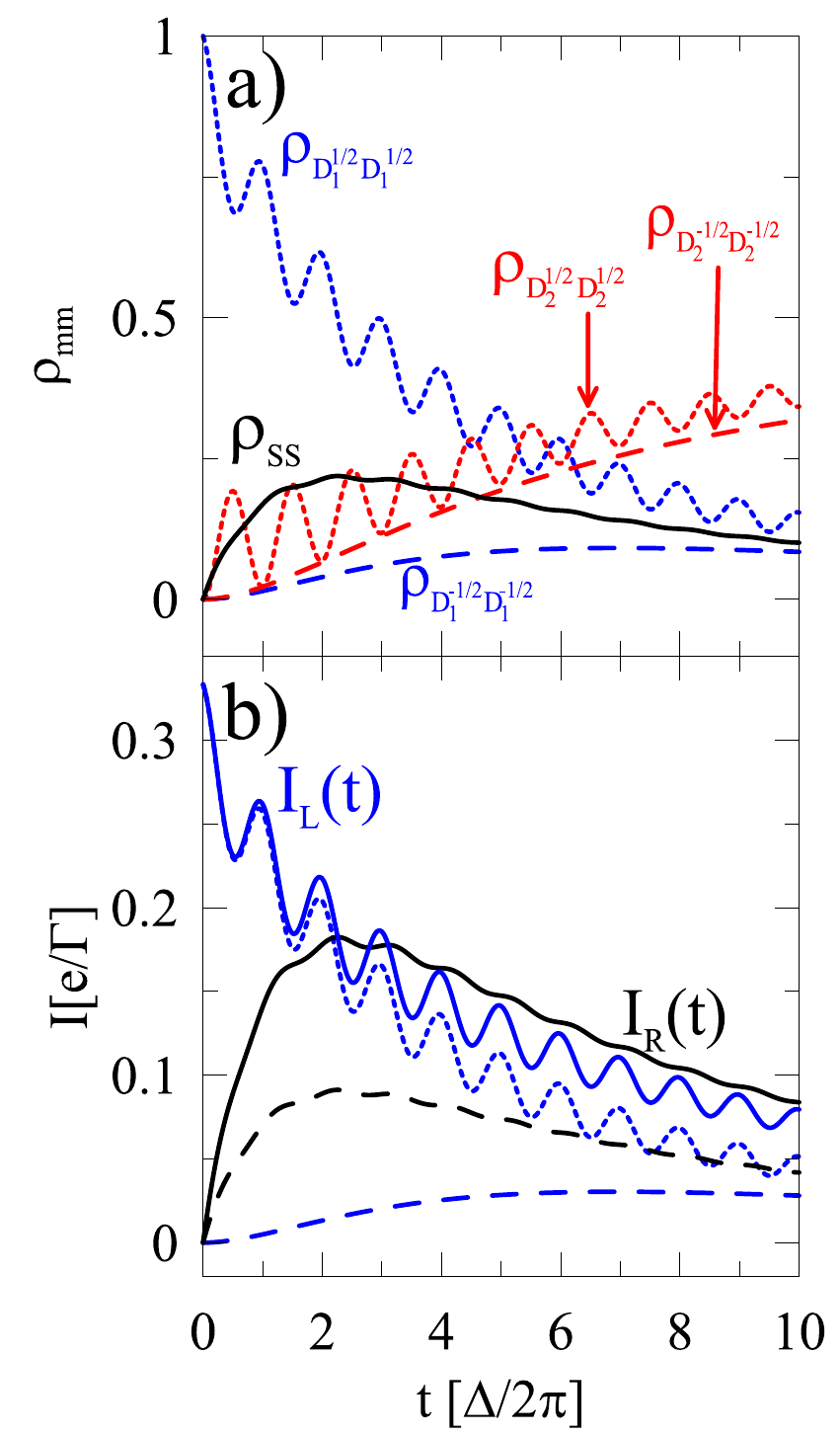}
\caption{Coherent oscillations including spin-flip processes when singlet in engaged in the current flow. Panel a) presents the density matrix elements for all
considered states; panel b) shows the time dependence of the current flowing through the left junction with spin $+1/2$ (blue dotted curve), $-1/2$ (blue dashed
curve) and right junction with spin $\pm1/2$ (black dashed curve). Notice that $I_R^{\uparrow}(t)=I_R^{\downarrow}(t)$. The total currents
$I_{L/R}(t)=I^{\uparrow}_{L/R}(t)+I^{\downarrow}_{L/R}(t)$ are plotted as a solid lines. The parameters taken in the calculations are the same as in Fig.
\ref{singdynamic}. The relaxation rates are
$1/T_{leak}=0.05$, $1/T_1=0.0069$, $1/T_2=0.0083$, $1/T^+_{sf}=0.0163$, and $1/T^-_{sf}=0.0003$.}
\label{spinflip}
\end{figure}

We make the Laplace transformation of the Master equation (\ref{master-spin-flip}) and find the relaxation rates from the poles of the polynomial
$z\,P^S(z)\,Q(z)$. Here the polynomial $P^S(z)$ is the same as for the previous case [described by Eq. (\ref{bloch-sing})] with the transfer rates including
degeneracy of the doublet states.
The second polynomial $Q(z)=z (z+\Gamma^{L-}_{D_1\to S}) +  \frac{1}{2}\gamma_0^2(z+\frac{1}{2}\Gamma^{L-}_{D_1\to S}) (2z +\Gamma^{L-}_{D_1\to
S})/\left[(z+\frac{1}{2}\Gamma^{L-}_{D_1\to S})^2+\delta^2\right]$ is related with the spin flip-processes which mix two doublet subspaces. From $Q(z)=0$ one
finds the spin-flip relaxation times
\begin{eqnarray}
\frac{1}{T_{sf}^{\pm}}=\left(1\pm\frac{|\delta|}{\Delta}\right)\,\frac{\Gamma^{L-}_{D_1\to S}}{2}
\end{eqnarray}
in the limit of weak coupling with the electrodes. The rate $1/T^+_{sf}$ is the second largest rate,
after leakage and describes a rapid relaxation process, which can be seen in the short time scale.
$1/T^-_{sf}$ corresponds the longest relaxation process which leads to total mixing of two doublet subspaces in the stationary limit. The dynamics in the doublet
subspace including spin-flip processes is presented in Fig.~\ref{spinflip}a.
The time evolution of \ket{D_1^{1/2}} is similar to the case without spin-flip (compare with Fig. \ref{singdynamic}), but now its reduction in the short time
scale  is faster due to $1/T_{sf}^+$. At the same time the \ket{D_1^{-1/2}} is built-up with the rate $1/T_{sf}^+$, and goes to the stationary limit
$\rho_{D_1^{1/2}D_1^{1/2}}(\infty)=\rho_{D_1^{-1/2}D_1^{-1/2}}(\infty)=0.03$ with the rate $1/T_{sf}^-$.

The qubit dynamics can be also seen in the spin current flowing through the left $I^{\sigma}_L(t)=e\Gamma^{L-}_{D_1\to
S}\rho_{D^{\sigma}_1D^{\sigma}_1}(t)$ and right junction $I^{\uparrow}_R(t)=I^{\downarrow}_R(t)=e(\Gamma^{R+}_{S\to D_1}+\Gamma^{R+}_{S\to
D_2})\rho_{SS}(t)$ -- see Fig.\ref{spinflip}~b.
The shape of the total current $I_{L}(t)=I^{\uparrow}_{L}(t)+I^{\downarrow}_{L}(t)$ is very similar as in the case without spin-flip. To get more information one
needs to measure the spin dependent currents $I^{\sigma}_L(t)$. The dashed and dotted blue curves in Fig.~\ref{spinflip}~b show $I^{\uparrow}_L(t)$ and
$I^{\downarrow}_L(t)$ with the characteristic times $T_{sf}^{\pm}$ in the short and long time scale. The fast increase of the total current in the right
electrode $I_R(t)$ for the very short time scale is related with $T_{leak}$ which now is two times shorter.

Let us estimate the characteristic times calculated in the paper. In the first order of approximation the relaxation times are proportional to $1/\Gamma^{L(R)+}_{\nu_2\rightarrow\nu_3}$ defined in equation (\ref{transferRates}). The tunnel rate $\Gamma$ in (\ref{transferRates}) is the order of neV for sequential transport \cite{thalakulam}. However it can be much larger in the coherent regime, $\Gamma\approx100$ $\mu$eV \cite{gores}. If we assume $\Gamma=1$ $\mu$eV for our system then the relaxation and decoherence time is $T_1\approx 4.8$ ns and $T_2 \approx 3.9$ ns for the case with singlet. For triplet we have $T_1 \approx 10$ ns and $T_2 \approx 6$ ns. The leakage to the singlet and triplet states are $1.1$ ns and $0.8$ ns respectively. These relaxation times are the same order as the decoherence time $T_2^*\sim 10$ ns due to hyperfine interaction in GaAs-based quantum dots \cite{laird2}. For the spin-flip processes the relaxation times are: $T_{sf}^+\approx 2$ ns and $T_{sf}^-\approx 109$ ns. One can see that due to the long relaxation time $T_{sf}^- $ the qubit conserves its spin coherence for a time needed for a read-out process.

\section{conclusion}

Summarizing we have proposed the qubit controlled by a symmetry breaking effect in a triangular TQD system. The main result of the paper is the new method for read-out of the qubit state by the current measurement in the doublet blockade regime, and the analysis of the qubit dynamics in the presence of decoherence processes caused by interaction with the electrodes.

We assumed that each dot contains one spin and the qubit was encoded in the doublet subspace. The qubit states has been controlled by the applied gate potentials which break the triangular symmetry. The calculations have been performed in the the Heisenberg model where the exchange couplings are modified by the orientation $\theta$ of the electric field with respect to the triangular axes.
For a specific $\theta$ one of the doublets is occupied and can be taken as an initial qubit state for further manipulations. By quick impulses of the electric field one can perform the Pauli X-gate and Z-gate operations. A composition of these two operations gives  full unitary control of the single qubit.

Moreover we have demonstrated the new method to read-out of the qubit states using the electric transport through TQD and the doublet blockade effect.
The method is compatible with pure electrical manipulations and the spin-to-charge conversion is not necessary.
The doublet blockade effect is related with an asymmetry of tunnel rates between the doublet states and the electrodes.
For some specific symmetry of TQD one of the doublet states is a dark one and the electron transport is blocked.
We have considered two cases with the singlet and the triplet as a ground state for two electrons. For the singlet case the current is blocked due to the doublet \ket{D_2}, whereas for transport from the triplet the dark state is the doublet \ket{D_1}. The doublet blockade can be also used to detect the qubit states in the linear TQD. However to satisfy the blockade condition $\gamma=0$ one of the electrodes must be connected to the central dot. Moreover the blockade can be applied to dynamical initialization of the qubit state as well as to perform Landau-Zener passages \cite{shevchenko}.

We have also considered the time dependent electron transport in the doublet blockade regime. Our
research gives information about dynamics of the qubit, the coherent oscillations and the relaxation processes due to presence of the electrodes. A role of the leakage
processes from the doublet to two electron states has been studied as well. For the triplet case the leakage is larger than for
singlet due to activation of the quadruplet state. We have also presented the driven case where the mixing parameter between the doublet state is
time dependent $\gamma(t)=\gamma_0 \exp(-i \omega t)$. In the resonance condition $\omega=\delta$ the doublet blockade is partially removed and one can observe strong Rabi oscillations.
Moreover we have investigated mixing of the doublet subspaces with $S_z=1/2$ and $S_z=-1/2$ caused by the spin-flip processes in the electrodes. The total mixing time $T_{sf}^- $ is very long what is promising for manipulation and read-out of the qubit.

\acknowledgments{We would like to thank Gloria Platero for discussion and valuable remarks. This work has been supported by the  National Science Centre under the contract DEC-2012/05/B/ST3/03208.}

\end{document}